\documentclass[11pt]{article}
\usepackage[all]{xy}
\usepackage{amssymb}
\usepackage{amsmath}
\usepackage{epstopdf}
\usepackage{graphicx}
\usepackage{amsthm}

\newtheorem{proposition}{Proposition}

\newtheorem{conjecture}{Conjecture}
\newtheorem{lemma}{Lemma}[]

\newtheorem{definition}{Definition}
\usepackage{authblk}
\usepackage[plain]{algorithm2e}
\date{}
\begin{document}

\title{\bf Algorithms for determining transposons in gene sequences}
\author[1,2,*]{Yue Wang}
\affil[1]{Department of Computational Medicine, University of California, Los Angeles, California, United States of America}
\affil[2]{School of Mathematical Sciences, Peking University, Beijing, China}
\affil[*]{E-mail address: yuew@g.ucla.edu. ORCID: 0000-0001-5918-7525}

\maketitle{}

\begin{abstract}
Some genes can change their relative locations in a genome. Thus for different individuals of the same species, the orders of genes might be different. Such jumping genes are called transposons. A practical problem is to determine transposons in given gene sequences. Through an intuitive rule, we transform the biological problem of determining transposons into a rigorous mathematical problem of determining the longest common subsequence. Depending on whether the gene sequence is linear (each sequence has a fixed head and tail) or circular (we can choose any gene as the head, and the previous one is the tail), and whether genes have multiple copies, we classify the problem of determining transposons into four scenarios: (1) linear sequences without duplicated genes; (2) circular sequences without duplicated genes; (3) linear sequences with duplicated genes; (4) circular sequences with duplicated genes. With the help of graph theory, we design fast algorithms for different scenarios. We also derive some results that might be of theoretical interests in combinatorics.
\begin{flushleft}
{\bf KEY WORDS:} transposon, gene sequence, algorithm, graph
\end{flushleft}
\end{abstract}

\section{Introduction}
Various genome rearrangement events, such as inversion, insertion, deletion, and duplication, can change the gene sequence. Such rearrangement events lead to the existence of transposons (also called transposable elements or jumping genes), which are DNA sequences that can change their relative positions within the genome of a cell. The mechanism of such transpositions can be expressed as a mixture of copy, cut, paste, and invert \cite{ivics2010expanding}. Transposons are common in various species. For the human genome, the proportion of transposons is approximately 44\% \cite{mills2007transposable}. Transposons can participate in controlling gene expression \cite{zhou2020dna}, and they are related to several diseases, such as cancer \cite{denicola2015utility}, hemophilia \cite{kazazian1988haemophilia}, and porphyria \cite{mustajoki1999insertion}. Transposons can drive rapid phenotypic variations, which cause complicated cell behaviors \cite{zhou2014multi,niu2015phenotypic,niu2019transposable,chen2016overshoot,jiang2017phenotypic}. Transposons can be used to detect cancer drivers \cite{noorani2020crispr} and potential therapies \cite{angelini2022model}. Transposons are also essential for the development of \emph{Oxytricha trifallax} \cite{nowacki2009functional}, antibiotic resistance of bacteria \cite{babakhani2018transposons}, and the proliferation of various cells \cite{rahrmann2009identification,xia2020pde,dessalles2021naive}. With the presence of transposons, the regulation between genes might be affected, which is a challenge for inferring the structures of gene regulatory networks \cite{wang2022inference} and general transcriptome analysis \cite{sha2020inference,zhou2021dissecting}.

There have been many algorithms developed to determine transposons, such as MELT \cite{gardner2017mobile}, ERVcaller \cite{chen2019ervcaller}, and TEMP2 \cite{yu2021benchmark}. For more details, readers may refer to other papers \cite{orozco2020measuring,goubert2022beginner}. However, they aim at targeting transposons (possibly very short, not whole genes) from raw DNA sequencing data. The sequencing data only contain imperfect information about the true DNA sequence, and the data quality depends on some factors that vary across different datasets \cite{evrony2021applications}. Besides, they need a corresponding genome or reference transposon libraries. Therefore, these algorithms focus more on the implementation aspect, not the theoretical aspect of the transposon determination problem.

In this paper, we consider an ideal scenario: We have some accurate sequences of genes (not nucleotides) from different individuals. Since some genes are transposons, these gene sequences are different. The goal is to compare these sequences and determine the transposons. 

In the copy-paste (duplication) case and deletion case, we can compare the numbers of copies of genes for different individuals to determine the transposons that have changed their copy numbers. In the inversion case, we can check the direction of genes to determine transposons that have changed their orientations \cite{lin2011changes}. In the cut-paste (insertion) case, the compositions of gene sequences are the same, but the orders of genes differ. It is not straightforward to uniquely determine which genes have changed their relative locations. Instead, we can consider the complement of transposons, which keep their relative locations and form a common subsequence of gene sequences from different individuals. Notice that genes in a subsequence does not need to be adjacent in the original sequences, different from a substring. We aim at explaining the difference among gene sequences with minimal transposons, meaning that we want to maximize the length of the complement of transposons. Thus we define the transposons to be the complement of the longest common subsequence.

It is common to use the length of the longest common subsequence as a quantitative score for comparing DNA sequences \cite{chen2004space,imbeault2017krab,zimin2017hybrid}. The longest common subsequence has also been used to define ultraconserved elements \cite{reneker2012long} or remove incongruent markers \cite{diop2020pseudomolecule}.

Determining the longest common subsequence is a classical problem in computer science. In the most commonly studied scenario, there are two sequences with possibly repeated genes, and the sequence length is $n$. The goal is to find the longest common subsequence, where the length is count by gene copies. This can be solved by dynamic programming with $\mathcal{O}(n^2)$ time complexity \cite{hirschberg1975linear}, but $\mathcal{O}(n^{2-\epsilon})$ time complexity for any $\epsilon>0$ is impossible \cite{backurs2015edit}. This also can be solved with $o(n)$ space complexity and $\mathcal{O}(n^3)$ time complexity \cite{kiyomi2021longest}. In another commonly studied scenario, there are $m$ sequences with possibly repeated genes, and the sequence length is $n$. The goal is to find the longest common subsequence, where the length is count by gene copies. This problem is equivalent to the maximum clique problem in graph theory, which is NP-hard \cite{maier1978complexity}. A standard dynamic programming algorithm has $\mathcal{O}(n^m)$ time complexity \cite{blum2021solving}. There have been other faster algorithms \cite{wang2010fast,mousavi2012improved,islam2019chemical}. For more works in these two classical scenarios, readers may refer to more thorough reviews \cite{bergroth2000survey,huang2004fast,wei2020path}.

In this paper, we consider four scenarios that are different from the classical ones. These four scenarios are determined by two factors: whether the gene sequence is linear or circular (since some species have circular DNAs while others not), and whether genes have multiple copies. When genes have multiple copies, we only consider common subsequences that consist of all or none of copies of the same gene. Scenario 1 has linear sequences without duplicated genes; Scenario 2 has circular sequences without duplicated genes; Scenario 3 has linear sequences with duplicated genes; Scenario 4 has circular sequences with duplicated genes. Scenarios 2,4 have circular sequences, and Scenarios 3,4 only consider subsequences that consist of all or none copies of the same gene, and calculate the length by genes. As far as we know, these new settings are not studied by other papers, and known methods cannot be directly applied. Thus we need to develop new algorithms, such as Algorithm~\ref{alg3} for Scenario 2. Besides, known methods only aim at finding one longest common subsequence. When the longest common subsequence is not unique, we also need to classify whether a gene appears in all/some/none of the longest common subsequences. Determining all longest common subsequences is too time-consuming, and we develop corresponding algorithms with polynomial time complexities for Scenarios 1,2 (Algorithms~\ref{alg2},\ref{alg4}). 

Scenario 1 (except the case with multiple longest common subsequences) is similar to well-studied classical situations, and our method (Algorithm~\ref{alg1}) is easily derived from standard algorithms. Scenarios 3,4 are equivalent to maximum clique problems in graphs and hypergraphs, which are NP-hard. These properties are also similar to the classical situations. For these NP-hard scenarios, we design fast heuristic algorithms (Algorithms~\ref{alg5},\ref{alg7}) and test them to find that they only fail in rare cases. These heuristic algorithms are similar to those classical algorithms for the maximum clique problem \cite{wu2015review}.

The author of this paper proposed the idea of using the longest common subsequence to find transposons and Algorithm~\ref{alg1} in a previous paper \cite{kang2014flexibility}, where other coauthors applied Algorithm~\ref{alg1} to study the ``core-gene-defined genome organizational framework'' (the complement of transposons) in various bacteria, and found that for different species, the transposon distribution and developmental traits are correlated. This paper considers other situations (especially when the longest common subsequence is not unique), and can be regarded as a theoretical sequel of that previous paper. Algorithm~\ref{alg1} is contained in this paper for the sake of completeness.

In sum, our main contributions are Algorithms~\ref{alg2},\ref{alg3},\ref{alg4} and Proposition~\ref{p1} that builds the equivalence between Scenario 3 and the maximum clique problem.

We first describe the setup for the problem of determining transposons and transform it into the problem of finding the longest common subsequence. In the following four scenarios, we transform them into corresponding graph theory problems and design algorithms. We finish with some discussions. All the algorithms in this paper have been implemented in Python. See https://github.com/YueWangMathbio/Transposon for the code files.

\section{Setup}
For some species, the DNA is a line \cite{rowley2018organizational}. We can represent this DNA as a linear gene sequence of distinct numbers that represent genes: $(1,2,3,4)$. If some genes change their transcriptional orientations, we can simply detect them and handle the remaining genes. Now a linear DNA naturally has a direction (from 5' end to 3' end), thus $(1,2,3,4)$ and $(4,3,2,1)$ are two different gene sequences. 

Consider two linear gene sequences from different individuals: $(1,2,3,4)$ and $(1,4,2,3)$. We can intuitively detect that gene $4$ changes its relative position, and should be regarded as a transposon. However, changing the positions of genes $2,3$ can also transform one sequence into the other. The reason that we think gene $4$ (not genes $2,3$) changes its relative position is that the number of genes we need to move is smaller. However, the number of genes that change their relative locations is difficult to determine. We can consider the complement of transposons, i.e., genes that do not change their relative positions. These fixed genes can be easily defined as the longest common subsequence of given gene sequences. Here a common subsequence consists of some genes (not necessarily adjacent, different from a substring) that keep their relative orders in the original sequences. \emph{Thus transposons are the complement of this longest common subsequence.} Notice that the longest common subsequence might not be unique. We classify genes by their relations with the longest common subsequence(s). The motivation of classifying transposons with respect to the intersection and union of longest common subsequences is similar to defining essential variables with Markov boundaries in causal inference \cite{wang2020causal}.

\begin{definition}
	A gene is a \textbf{proper-transposon} if it is not contained in any longest common subsequence. A gene is a \textbf{non-transposon} if it is contained in every longest common subsequence. A gene is a \textbf{quasi-transposon} if it is contained in some but not all longest common subsequences.
	\label{def1}
\end{definition}

In the example of $(1,2,3,4)$ and $(1,4,2,3)$, the unique longest common subsequence is $(1,2,3)$. Thus $4$ is a proper-transposon, and $1,2,3$ are non-transposons. In the following, we consider other scenarios, where the proper/quasi/non-transposons still follow Definition~\ref{def1}, but the definition of the longest common subsequence differs.

For some species, the DNA is a circle, not a line \cite{verma2019architecture}. A circular DNA also has a natural direction (from 5' end to 3' end), and we use the clockwise direction to represent this natural direction. In the circular sequence scenario, a common subsequence is a circular sequence that can be obtained from each circular gene sequence by deleting some genes. See Fig.~\ref{sce2} for two circular gene sequences and their longest common subsequence. Notice that we can rotate each circular sequence for a better match.

\begin{figure}[htb]
	\begin{center}
		$\xymatrix{
		1\ar@{-}[r]\ar@{-}[d]&2\ar@{-}[r]	& 3\ar@{-}[d] &3 \ar@{-}[r]\ar@{-}[d]&1\ar@{-}[r]&2\ar@{-}[d]&1\ar@{-}[r]\ar@{-}[d]&2\ar@{-}[d]\\
		6&	5\ar@{-}[l] & 4\ar@{-}[l]    &5 &4\ar@{-}[l]  &6   \ar@{-}[l]  &5&4\ar@{-}[l] 
		}$
	\end{center}
	\caption{Two circular gene sequences without duplicated genes and their longest common subsequence, corresponding to Scenario 2.}
	\label{sce2}
\end{figure}

A gene might have multiple copies (duplicated) in a gene sequence \cite{ibal2019information}. Notice that the definition of the transposon is a gene (specific DNA sequence) that has the ability to change its position, not a certain copy of a gene that changes its position. This means transposons should be defined for genes, not gene copies. Thus we should only consider common subsequences that consist of all or none copies of the same gene. When calculating the length of a common subsequence, we should count genes, not gene copies. Consider two linear sequences $(4,1,2,1,1,3,2,4,1,1)$ and $(4,1,2,3,1,1,2,1,1,4)$. If we consider any subsequences, the longest common subsequence is $(4,1,2,1,1,2,1,1)$; if we only consider subsequences that contain all or none copies of the same gene, but count the length by copies, the longest common subsequence is $(1,2,1,1,2,1,1)$; if we only consider subsequences that contain all or none copies of the same gene, and count the length by genes, the unique longest common subsequence is $(4,2,3,2,4)$, and gene $1$ is a proper-transposon.

When we consider circular gene sequences with duplicated genes, we should still only consider subsequences that consist of all or none copies of the same gene, and calculate the length by genes. Notice that circular sequences can be rotated. See Fig.~\ref{sce4} for two circular gene sequences with duplicated genes and their longest common subsequence.

\begin{figure}[htb]
	\begin{center}
		$\xymatrix{
			1\ar@{-}[r]\ar@{-}[d]&2\ar@{-}[r]	& 1\ar@{-}[d] &3 \ar@{-}[r]\ar@{-}[d]&1\ar@{-}[r]&3\ar@{-}[d]&1\ar@{-}[r]\ar@{-}[d]&2\ar@{-}[d]\\
			3&	2\ar@{-}[l] & 3\ar@{-}[l]    &2 &1\ar@{-}[l]  &2   \ar@{-}[l]  &2&1\ar@{-}[l] 
		}$
	\end{center}
	\caption{Two circular gene sequences with duplicated genes and their longest common subsequence, corresponding to Scenario 4.}
	\label{sce4}
\end{figure}

We have turned the problem of determining transposons into finding the longest common subsequence of several gene sequences. Depending on whether the gene sequences are linear or circular, and whether genes have multiple copies, the problem can be classified into four scenarios:

\noindent \textbf{Scenario 1}: Consider $m$ linear sequences of genes $1,\ldots,n$, where each gene has only one copy in each sequence. Determine the longest linear sequence that is a common subsequence of these $m$ sequences.

\noindent \textbf{Scenario 2}: Consider $m$ circular sequences of genes $1,\ldots,n$, where each gene has only one copy in each sequence. Determine the longest circular sequence that is a common subsequence of these $m$ sequences. Here circular sequences can be rotated.

\noindent \textbf{Scenario 3}: Consider $m$ linear sequences of genes $1,\ldots,n$, where each gene can have multiple copies in each sequence. Determine the longest linear sequence that is a common subsequence of these $m$ sequences. Only consider subsequences that consist of all or none copies of the same gene, and calculate the length by genes.

\noindent \textbf{Scenario 4}: Consider $m$ circular sequences of genes $1,\ldots,n$, where each gene can have multiple copies in each sequence. Determine the longest circular sequence that is a common subsequence of these $m$ sequences. Only consider subsequences that consist of all or none copies of the same gene, and calculate the length by genes. Here circular sequences can be rotated.

These four scenarios correspond to different algorithms, and will be discussed separately.

\section{Linear sequences without duplicated genes}

In Scenario 1, consider $m$ linear gene sequences, where each sequence contains $n$ genes $1,\ldots,n$. Each gene has only one copy. For such permutations of $1,\ldots,n$, we need to find the longest common subsequence. 

\subsection{A graph representation of the problem}
Brute-force searching that tests whether each subsequence appears in all sequences is not applicable, since the time complexity is exponential in $n$. To develop a polynomial algorithm, we first design an auxiliary directed graph $\mathcal{G}$.

\begin{definition}
	For $m$ linear sequences with $n$ non-duplicated genes, the corresponding \textbf{auxiliary graph} $\mathcal{G}$ is a directed graph, where each vertex is a gene $g_i$, and there is a directed edge from $g_i$ to $g_j$ if and only if $g_i$ appears before $g_j$ in all $m$ sequences.
\end{definition}

A directed path $g_1\to g_2\to g_3\to\cdots\to g_4\to g_5$ in $\mathcal{G}$ corresponds to a common subsequence $(g_1,g_2,g_3,\ldots,g_4,g_5)$ of $m$ sequences, and vice versa. We add $0$ to the head of each sequence and $n+1$ to the tail. Then the longest common subsequence must start at $0$ and end at $n+1$. \emph{The problem of finding the longest common subsequence becomes finding the longest path from $0$ to $n+1$ in $\mathcal{G}$.} See Fig.~\ref{ag} for an example of using the auxiliary graph to determine transposons.  This auxiliary graph $\mathcal{G}$ has no directed loop (acyclic). If there exists a loop $g_1\to g_2\to g_3\to\cdots\to g_4\to g_1$, then $g_1$ is prior to $g_4$ and $g_4$ is prior to $g_1$ in all sequences, a contradiction.

\begin{figure}[htb]
	\begin{center}
		$\xymatrix{
		& 0\ar@2{->}[dl]\ar[dr]\ar[ddl]\ar[ddr]\ar[ddd] &\\
		1\ar[d]\ar@2{->}[rr]\ar[drr]\ar[ddr] &     & 2 \ar@2{->}[dll]\ar[ddl]  \\
		3\ar@2{->}[dr]     &     & 4\ar[dl]      \\
		&5&         }$
	\end{center}
	\caption{The auxiliary graph $\mathcal{G}$ of two sequences $([0],1,2,3,4,[5])$ and $([0],1,4,2,3,[5])$. The unique longest path (double arrows) from $0$ to $5$ is $0\to 1\to 2\to 3\to 5$, meaning that the unique longest common sequence is $([0],1,2,3,[5])$. Thus $1,2,3$ are non-transposons, and $4$ is a proper-transposon.}
	\label{ag}
\end{figure}

\subsection{Find the longest path}
Determining the longest path between two vertices in a directed acyclic graph can be solved by a standard dynamic programming algorithm. For a vertex $g_i\in\{0,1,\ldots,n\}$, consider the longest path from $g_i$ to $n+1$. Since there exists an edge $g_i\to n+1$, and $\mathcal{G}$ is acyclic, this longest path exists. If the longest path is not unique, assign one arbitrarily. 
\begin{definition}
Define $F_+(g_i)$ to be the length of the longest path from $g_i$ to $n+1$ in $\mathcal{G}$, and $H_+(g_i)$ to be the vertex next to $g_i$ in this path. 
\end{definition}
$F_+$ and $H_+$ can be calculated recursively: For one gene $g_i$, consider all genes $g_j$ with an edge $g_i\to g_j$ in $\mathcal{G}$. The gene $g_j$ with the largest $F_+(g_j)$ is assigned to be $H_+(g_i)$, and $F_+(g_i)=F_+(g_j)+1$. If $g_l\to n+1$ is the only edge that starts from gene $g_l$, then $F_+(g_l)=1$, and $H_+(g_l)=n+1$. In other words,
\[H_+(g_i)=\underset{\{g_j\text{ with }g_i\to g_j\}}{\mathrm{argmax}}\,F_+(g_j);\]
\[F_+(g_i)=1+F_+[H_+(g_i)].\]
Then $0\to H_+(0)\to H_+^{2}(0)\to H_+^{3}(0)\to 
\cdots{}\to H_+^{f-1}(0)\to H_+^{f}(0)=n+1$, denoted by $\mathcal{L}_0$, is a longest path in $\mathcal{G}$. Here $f=F_+(0)$, and $H_+^i$ is the $i$th iteration of $H_+$. 

\subsection{Test the uniqueness of the longest path}
To test whether quasi-transposons exist, we need to check the uniqueness of this longest path.
\begin{definition}
	For $g_i\in\{1,\ldots,n, n+1\}$, define $F_-(g_i)$ to be the length of the longest path from $0$ to $g_i$ in $\mathcal{G}$, and $H_-(g_i)$ to be the vertex prior to $g_i$ in this path. 
\end{definition}
$F_-$ and $H_-$ can be calculated similar to $F_+$ and $H_+$. We can see that $F_+(g_i)+F_-(g_i)$ is the length of 
\[0=H_-^{F_-(g_i)}(g_i)\to H_-^{F_-(g_i)-1}(g_i) \to \cdots \to H_-(g_i) \to g_i\]
\[\to H_+(g_i) \to \cdots\to H_+^{F_+(g_i)-1}(g_i)\to H_+^{F_+(g_i)}(g_i)=n+1,\]
a longest path from $0$ through $g_i$ to $n+1$. For $g_i\notin \mathcal{L}_0$, if $F_+(g_i)+F_-(g_i)<F_+(0)$, then $g_i$ is a proper-transposon; if $F_+(g_i)+F_-(g_i)=F_+(0)$, then $g_i$ is a quasi-transposon. If every $g_i\notin \mathcal{L}_0$ is a proper-transposon, then the longest common subsequence is unique, and all genes in $\mathcal{L}_0$ (excluding the auxiliary $0$ and $n+1$) are non-transposons. The procedure of determining transposons stops here. Otherwise, the longest common subsequence is not unique, and we need to find quasi-transposons in $\mathcal{L}_0$.

\subsection{Find quasi-transposons}
When determining all quasi-transposons $g_1,\ldots,g_k$ not in $\mathcal{L}_0$, as described above, we construct corresponding longest paths $\mathcal{L}_1,\ldots,\mathcal{L}_k$ from $0$ to $n+1$, where each $\mathcal{L}_i$ passes through $g_i$. We claim that a gene $g_j\in \mathcal{L}_0$ is a non-transposon if and only if $g_j$ is contained in all $\mathcal{L}_1,\ldots,\mathcal{L}_k$. To prove this, we need the following lemma.

\begin{lemma}
	In Scenario 1 of linear sequences without duplicated genes, each quasi-transposon $g_i$ has a corresponding quasi-transposon $g_j$, so that no longest common subsequence can contain both $g_i$ and $g_j$.
	\label{l1}
\end{lemma}

If a gene $g_j\in \mathcal{L}_0$ is a non-transposon, then it is contained in all $\mathcal{L}_1,\ldots,\mathcal{L}_k$. If $g_j\in \mathcal{L}_0$ is a quasi-transposon, by Lemma~\ref{l1}, there is a quasi-transposon $g_l\notin\mathcal{L}_0$ which is mutual-exclusive with $g_j$, in the sense that $g_l$ and $g_j$ cannot appear in the same longest common subsequence. The corresponding longest path $\mathcal{L}_l$ contains $g_l$, thus cannot contain $g_j$. This proves our approach to determine the quasi-transposons in $\mathcal{L}_0$.

\begin{proof}[Proof of Lemma~\ref{l1}]
	Fix a quasi-transposon $g_i$. It is contained in a longest path $\mathcal{L}_i$, which contains all non-transposons. Thus for each non-transposon $g^*$, there is an edge between $g^*$ and $g_i$ in $\mathcal{G}$. Assume $g_i$ has no such mutual-exclusive quasi-transposon $g_j$. Then there is an edge (direction unknown) in $\mathcal{G}$ between $g_i$ and each quasi-transposon $g_j$. Choose a longest path $\mathcal{L}^*$ in $\mathcal{G}$ that does not contain $g_i$. Whether $g_j\in\mathcal{L}^*$ is a non-transposon or a quasi-transposon, there is an edge between $g_j$ and $g_i$. Determine the first gene $g_k$ in $\mathcal{L}^*$ that has an edge $g_i\to g_k$. Since there is an edge $g_i\to n+1$, $g_k$ exists. Since there is an edge $0\to g_i$, $g_k\ne 0$. Denote the previous gene of $g_k$ in $\mathcal{L}^*$ by $g_l$, then $g_l$ exists, and there is an edge $g_l\to g_i$. Thus we construct a path $0\to\cdots\to g_l\to g_i\to g_k\to \cdots\to n+1$, which is longer than the longest path, a contradiction. Thus $g_i$ has a mutual-exclusive quasi-transposon $g_j$.
\end{proof}

\subsection{Algorithms and complexities}

\begin{algorithm}[!htbp]
	\caption{Detailed workflow of determining proper-transposons and quasi-transposons in Scenario 1, preparation stage.}
	\label{alg1}
	\vspace{-\bigskipamount}
	\ \\
	\begin{enumerate}
		{	\item \textbf{Input} 
			
			\quad $m$ linear sequences of genes $1,\ldots,n$. No duplicated genes.
			
			\item \textbf{Modify} the sequences:
			
			\quad Add $0$ to the head, and $n+1$ to the tail of each sequence 
			
			\item \textbf{Construct} the auxiliary graph $\mathcal{G}$:
			
			\quad Vertices of $\mathcal{G}$ are all the genes $1,\ldots,n$
			
			\quad \textbf{For} each pair of genes $g_i,g_j$
			
			\quad\quad   \textbf{If} $g_i$ is prior to $g_j$ in all $m$ sequences
			
			\quad\quad\quad \textbf{Add} a directed edge $g_i\to g_j$ in $\mathcal{G}$
			
			\quad\quad\textbf{End} of if
			
			\quad\textbf{End} of for
			
			\item \textbf{Calculate} $F_+(\cdot)$ and $H_+(\cdot)$ for each gene $g_i$ in $0,1,\ldots,n$ recursively; \textbf{calculate} $F_-(\cdot)$ and $H_-(\cdot)$ for each gene $g_i$ in $1,\ldots,n, n+1$ recursively: 
			
			\quad $H_+(g_i)=\underset{\{g_j\text{ with }g_i\to g_j\}}{\mathrm{argmax}}\,F_+(g_j)$ 
			
			\quad \% If $g_j$ with $g_i\to g_j$ that maximizes $F_+(g_j)$ is not unique, choose one randomly
			
			\quad $F_+(g_i)=1+F_+[H_+(g_i)]$ 
			
			\quad $H_-(g_i)=\underset{\{g_j\text{ with }g_j\to g_i\}}{\mathrm{argmax}}\,F_-(g_j)$
			
			\quad \% If argmax is not unique, choose one randomly
			
			\quad $F_-(g_i)=1+F_-[H_-(g_i)]$
			
			\item \textbf{Construct} a longest path $\mathcal{L}_0$ from $0$ to $n+1$:
			
			\quad $0\to H_+(0)\to H_+^{2}(0)\to H_+^{3}(0)\to 
			\cdots{}\to H_+^{f-1}(0)\to H_+^{f}(0)=n+1$
			
			\quad \% Here $f=F_+(0)$, and $H_+^i$ is the $i$th iteration of $H_+$
			
			\item \textbf{Output} $F_+(\cdot),H_+(\cdot),F_-(\cdot),H_-(\cdot),\mathcal{L}_0$
		}
	\end{enumerate}
\end{algorithm}

\begin{algorithm}[!htbp]
	\caption{Detailed workflow of determining proper-transposons and quasi-transposons in Scenario 1, output stage.}
	\label{alg2}
	\vspace{-\bigskipamount}
	\ \\
	\begin{enumerate}
		{	\item \textbf{Input} 
			
			\quad $F_+(\cdot),H_+(\cdot),F_-(\cdot),H_-(\cdot),\mathcal{L}_0$ calculated from Algorithm~\ref{alg1}				
			
			\quad \textbf{Denote} all genes not in $\mathcal{L}_0$ by $g_1,\ldots,g_k$
			
			\item \textbf{For} each gene $g_i$ in $g_1,\ldots,g_k$
			
			\quad \textbf{If} $F_+(g_i)+F_-(g_i)<F_+(0)$
			
			\quad\quad \textbf{Output} $g_i$ is a proper-transposon
			
			\quad \textbf{Else}
			
			\quad\quad \textbf{Output} $g_i$ is a quasi-transposon
			
			\quad \textbf{End} of if
			
			\textbf{End} of for
			
			\item \textbf{If} all genes in $g_1,\ldots,g_k$ are proper-transposons
			
			\quad \textbf{Output} all genes in $\mathcal{L}_0$ are non-transposons
			
			\textbf{Else} 
			
			\quad \textbf{For} each gene $g_i$ in $g_1,\ldots,g_k$ 
			
			\quad\quad Use $H_+(\cdot)$ and $H_-(\cdot)$ to \textbf{construct} $\mathcal{L}_i$, a longest path from $0$ to $n+1$ that passes $g_i$.
			
			\quad \textbf{End} of for
			
			\quad \textbf{For} each gene $g_j$ in $\mathcal{L}_0$ (excluding auxiliary $0$ and $n+1$)
			
			\quad\quad \textbf{If} $g_j$ is contained in all $\mathcal{L}_1,\ldots,\mathcal{L}_k$
			
			\quad\quad\quad \textbf{Output} $g_j$ is a non-transposon
			
			\quad\quad \textbf{Else}
			
			\quad\quad\quad \textbf{Output} $g_j$ is a quasi-transposon
			
			\quad\quad \textbf{End} of if
			
			\quad\textbf{End} of for
			
			\textbf{End} of if
			
			\item\textbf{Output}: whether each gene is a proper/quasi/non-transposon
			
		}
	\end{enumerate}
\end{algorithm}

We summarize the above method as Algorithms~\ref{alg1},\ref{alg2}. If we have known that the longest common subsequence is unique, then we just need to apply Algorithm~\ref{alg1}, so that genes in $\mathcal{L}_0$ are non-transposons, and genes not in $\mathcal{L}_0$ are proper-transposons. Algorithm~\ref{alg1} has been briefly reported in a previous paper, also by the author of this paper \cite{kang2014flexibility,wang2018some}. We keep Algorithm~\ref{alg1} here to make the story complete. Assume we have $m$ sequences with length $n$, and the length of the longest common subsequence is $n-k$. The time complexities of Steps 2-5 in Algorithm~\ref{alg1} are $\mathcal{O}(m)$, $\mathcal{O}(mn^2)$, $\mathcal{O}(n)$, $\mathcal{O}(n)$. The time complexities of Step 2 and Step 3 in Algorithm~\ref{alg2} are $\mathcal{O}(k)$ and $\mathcal{O}(kn)$. Since $k\le n$, the overall time complexity of determining transposons in Scenario 1 by Algorithms~\ref{alg1},\ref{alg2} is $\mathcal{O}(mn^2)$. The space complexity is trivially $\mathcal{O}(mn+n^2)$.

\section{Circular sequences without duplicated genes}

In Scenario 2, consider $m$ circular gene sequences, where each sequence contains $n$ genes $1,\ldots,n$. Each gene has only one copy in each sequence. For such circular permutations of $1,\ldots,n$, we need to find the longest common subsequence. Assume the length of the longest common subsequence is $n-k$.

\subsection{Find a longest common subsequence}
We first randomly choose a gene $g_i$. Cut all circular sequences at $g_i$ and expand them to be linear sequences. For example, the circular sequences in Fig.~\ref{sce2} cut at $1$ are correspondingly $(1,2,3,4,5,6)$ and $(1,2,6,4,5,3)$. Using Algorithm~\ref{alg1}, we can find $\mathcal{L}_i$ that begins with $g_i$, which is a longest common subsequence of all expanded linear sequences. In the above example, the longest common linear subsequence starting from $1$ is $(1,2,4,5)$. If $g_i$ is a non-transposon or a quasi-transposon, then $\mathcal{L}_i$ (glued back to a circle) is a longest common circular subsequence. If $g_i$ is a proper-transposon, then $\mathcal{L}_i$ is shorter than the longest common circular subsequence. In Fig.~\ref{sce2}, gene $1$ is a non-transposon, and $(1,2,4,5)$ (glued) is the longest common circular subsequence.

We do not know if $\mathcal{L}_i$ (glued) is a longest common subsequence (whether containing $g_i$ or not) for all circular sequences. If there is a longer common subsequence, it should contain genes that are not in $\mathcal{L}_i$. Consider four variables $\mathcal{L}$, $g$, $C$, and $\mathcal{S}$, whose initial values are $\mathcal{L}_i$, $g_i$, the length of $\mathcal{L}_i$, and the complement of $\mathcal{L}_i$. These variables contain information on the longest common linear subsequence that we have found during this procedure.

Choose a gene $g_j$ in $\mathcal{S}$, and cut all circular gene sequences at $g_j$. Apply Algorithm~\ref{alg1} to find $\mathcal{L}_j$, which is the longest in common subsequences that contain $g_j$. If the length of $\mathcal{L}_j$ is larger than $C$, set $\mathcal{L}$ to be $\mathcal{L}_j$, set $g$ to be $g_j$, set $C$ to be the length of $\mathcal{L}_j$, and set $\mathcal{S}$ to be the complement of $\mathcal{L}_j$. Otherwise, keep $\mathcal{L}$, $g$, $C$, and $\mathcal{S}$ still.

Choose another gene $g_l$ in $\mathcal{S}$ which has not been chosen before, and repeat this procedure. This procedure terminates when all genes in $\mathcal{S}$ have been chosen and cut. Denote the final values of $\mathcal{L}$, $g$, $C$, and $\mathcal{S}$ by $\mathcal{L}_0$, $g_0$, $C_0$, and $\mathcal{S}_0$. Here $\mathcal{S}_0$ is the complement of $\mathcal{L}_0$.

During this procedure, if the current $g$ is a proper-transposon, then $\mathcal{S}$ contains a non-transposon or a quasi-transposon, which has not been chosen. Thus $\mathcal{L}$, $g$, $C$, $\mathcal{S}$ will be further updated. If the current $g$ is a non-transposon or a quasi-transposon, then $C$ has reached its maximum, and $\mathcal{L}$, $g$, $C$, $\mathcal{S}$ will not be further updated. This means $\mathcal{L}_0$ is a longest common circular subsequence, and $C_0$ is the length of the longest common subsequence, $n-k$. Also, the total number of genes being chosen and cut is $k+1$. All $k$ genes in $\mathcal{S}_0$ and $g_0$ are chosen and cut. A gene $g_t$ in $\mathcal{L}_0$ (excluding $g_0$) is a non-transposon or a quasi-transposon, and cannot be chosen and cut. The reason is that it cannot be chosen before $g_0$ is chosen (only proper-transposons can be chosen before $g_0$ is chosen), and it cannot be chosen after $g_0$ is chosen ($g_t\notin \mathcal{S}_0$).   


\subsection{Determine quasi-transposons}
For each gene $g_p\in\mathcal{S}_0$, apply Algorithm~\ref{alg1} to calculate $C_p$, the length of the longest common subsequence that contains $g_p$. If $C_p<C_0$, $g_p$ is a proper-transposon. Otherwise, $C_p=C_0$ means $g_p$ is a quasi-transposon. We have found all proper-transposons. If all genes in $\mathcal{S}_0$ are proper-transposons, then all genes in $\mathcal{L}_0$ are non-transposons, and the procedure terminates. 

If $\mathcal{S}_0$ contains quasi-transposons, then $\mathcal{L}_0$ also has quasi-transposons. To determine quasi-transposons in $\mathcal{L}_0$, we need the following lemma.
\begin{lemma}
In Scenario 2, choose a quasi-transposon $g_p$ and cut the circular sequences at $g_p$ to obtain linear sequences. A proper-transposon for the circular sequences is also a proper-transposon for the linear sequences; a non-transposon for the circular sequences is also a non-transposon for the linear sequences. 
\label{nl}
\end{lemma}
\begin{proof}
Consider a longest common subsequence $\mathcal{L}_p$ for linear sequences cut at $g_p$. Since $g_p$ is a quasi-transposon, the length of $\mathcal{L}_p$ is also $n-k$, meaning that $\mathcal{L}_p$ is also a longest common subsequence for circular sequences. Now, this lemma is proved by the definition of proper/quasi/non-transposon.
\end{proof}

If a gene $g_r$ in $\mathcal{L}_0$ is a non-transposon for the circular sequences, then $g_r$ is a non-transposon for linear sequences cut at each quasi-transposon $g_q\in\mathcal{S}_0$. If a gene $g_s$ in $\mathcal{L}_0$ is a quasi-transposon for the circular sequences, then there is a longest common circular subsequence $\mathcal{L}_t$ that does not contain $g_s$, meaning that $\mathcal{L}_t$ contains a quasi-transposon $g_t$ not in $\mathcal{L}_0$. Then $g_s$ is a proper/quasi-transposon for linear sequences cut at $g_t$. 

Therefore, we can use the following method to determine quasi-transposons in $\mathcal{L}_0$. For each quasi-transposon $g_q\in\mathcal{S}_0$, cut at $g_q$ and apply Algorithms~\ref{alg1},\ref{alg2} to determine if each gene in $\mathcal{L}_0$ is a proper/quasi/non-transposon for the linear gene sequences cut at $g_q$. A gene $g_r\in\mathcal{L}_0$ is a non-transposon for the circular sequences if and only if it is a non-transposon for linear sequences cut at any quasi-transposon $g_q\in\mathcal{S}_0$. A gene $g_s\in\mathcal{L}_0$ is a quasi-transposon for the circular sequences if and only if it is a proper/quasi-transposon for linear sequences cut at some quasi-transposon $g_q\in\mathcal{S}_0$.

When we have determined all quasi-transposons in $\mathcal{S}_0$, it might be tempting to apply a simpler approach to determine quasi-transposons in $\mathcal{L}_0$: For each quasi-transposon $g_q\in\mathcal{S}_0$, cut at $g_q$ and apply Algorithm~\ref{alg1} to find a longest common subsequence $\mathcal{L}_q$. A gene in $\mathcal{L}_0$ is a non-transposon if and only if it appears in all such $\mathcal{L}_q$. This approach is valid only if the following conjecture holds, which is similar to Lemma~\ref{l1}:

\begin{conjecture}
	In Scenario 2 of circular sequences without duplicated genes, each quasi-transposon $g_i$ has a corresponding quasi-transposon $g_j$, so that no longest common subsequence can contain both $g_i$ and $g_j$.
	\label{c1}
\end{conjecture}

However, Conjecture~\ref{c1} does not hold. See Fig.~\ref{ce} for a counterexample. All genes are quasi-transposons. Any two quasi-transposons are contained in a longest common subsequence (length $3$). Thus the simplified approach above does not work.

\begin{figure}[htb]
	\begin{center}
		$\xymatrix{
			1\ar@{-}[r]\ar@{-}[d]&2\ar@{-}[r]	& 3\ar@{-}[d] &1 \ar@{-}[r]\ar@{-}[d]&2\ar@{-}[r]&6\ar@{-}[d]&1\ar@{-}[r]\ar@{-}[d]&2\ar@{-}[r]&7\ar@{-}[d]\\
			8\ar@{-}[d]&	 & 4\ar@{-}[d]    &3\ar@{-}[d] & &5   \ar@{-}[d]  &6\ar@{-}[d]&&8\ar@{-}[d]\\
			7\ar@{-}[r]&6\ar@{-}[r]&5&4\ar@{-}[r]&7\ar@{-}[r]&8&5\ar@{-}[r]&4\ar@{-}[r]&3
		}$
	\end{center}
	\caption{A counterexample with three circular sequences that fails Conjecture~\ref{c1}.}
	\label{ce}
\end{figure}

We summarize the above method as Algorithms~\ref{alg3},\ref{alg4}. If we have known that the longest common subsequence is unique, then we just need to apply Algorithm~\ref{alg3}, so that genes in $\mathcal{S}_0$ are proper-transposons, and genes not in $\mathcal{S}_0$ are non-transposons. Assume we have $m$ sequences with length $n$, and the length of the longest common subsequence is $n-k$. The time complexities of Step 2 and Step 3 in Algorithm~\ref{alg3} are $\mathcal{O}(mn^2)$ and $\mathcal{O}(kmn^2)$. The time complexities of Step 2 in Algorithm~\ref{alg4} is $\mathcal{O}(kmn^2)$. The overall time complexity of determining transposons in Scenario 2 by Algorithms~\ref{alg3},\ref{alg4} is $\mathcal{O}(kmn^2)$. The space complexity is trivially $\mathcal{O}(mn+n^2)$. 

\begin{algorithm}[!htbp]
	\caption{Detailed workflow of determining proper-transposons and quasi-transposons in Scenario 2, preparation stage.}
	\label{alg3}
	\vspace{-\bigskipamount}
	\ \\
	\begin{enumerate}
		{	\item \textbf{Input}
			
			\quad $m$ circular sequences of genes $1,\ldots,n$, where each gene has only one copy in each sequence
			
			\item \textbf{Choose} a gene $g_i$ randomly 
			
			\textbf{Cut} all circular sequences at $g_i$ and expand them to be linear sequences
			
			\textbf{Apply} Algorithm~\ref{alg1} to find $\mathcal{L}_i$, a longest common subsequence in the expanded linear sequences
			
			\textbf{Set} $C$ to be the length of $\mathcal{L}_i$, and \textbf{set} $\mathcal{S}$ to be the complement of $\mathcal{L}_i$
			
			\item \textbf{While} $\mathcal{S}$ has a gene $g_j$ that has not been chosen and cut
			
			\quad \textbf{Cut} all circular sequences at $g_j$ and apply Algorithm~\ref{alg1} to find $\mathcal{L}_j$
			
			\quad \textbf{Denote} the length of $\mathcal{L}_j$ by $C_j$
			
			\quad \textbf{If} $C_j>C$
			
			\quad\quad   \textbf{Update} $C$ to be $C_j$, and \textbf{update} $\mathcal{S}$ to be the complement of $\mathcal{L}_j$
			
			\quad\textbf{End} of if
			
			\textbf{End} of while
			
			\textbf{Denote} the final $C$ by $C_0$, and \textbf{denote} the final $\mathcal{S}$ by $\mathcal{S}_0$
									
			\item \textbf{Output} $C_0$ and $\mathcal{S}_0$
		}
	\end{enumerate}
\end{algorithm}

\begin{algorithm}[!htbp]
	\caption{Detailed workflow of determining proper-transposons and quasi-transposons in Scenario 2, output stage.}
	\label{alg4}
	\vspace{-\bigskipamount}
	\ \\
	\begin{enumerate}
		{	\item \textbf{Input}
			
			\quad $m$ circular sequences of genes $1,\ldots,n$, where each gene has only one copy in each sequence; $C_0$ and $\mathcal{S}_0$ calculated from Algorithm~\ref{alg3}

			\item \textbf{For} each gene $g_l\in \mathcal{S}_0$ 	
			
			\quad \textbf{Cut} all circular sequences at $g_l$ and expand them to be linear sequences
			
			\quad\textbf{Apply} Algorithm~\ref{alg1} to find $\mathcal{L}_l$, a longest common subsequence in the expanded linear sequences.
			
			\quad \textbf{Denote} the length of $\mathcal{L}_l$ by $C_l$	
			
			\quad \textbf{If} $C_l<C_0$
			
			\quad\quad \textbf{Output} $g_l$ is a proper-transposon
			
			\quad \textbf{Else}
			
			\quad\quad \textbf{Output} $g_l$ is a quasi-transposon
			
			\quad\quad \textbf{Cut} all circular sequences at $g_l$ and \textbf{apply} Algorithms~\ref{alg1},\ref{alg2} to find all proper/quasi-transposons for linear gene sequences starting at $g_l$
			
			\quad\quad \textbf{Output} genes not in $\mathcal{S}_0$ but being proper/quasi-transposons for such linear sequences are quasi-transposons for circular sequences
			
			\quad \textbf{End} of if
			
			\textbf{End} of for		
			
			\textbf{Output} other genes that have not been determined to be proper/quasi-transposons are all non-transposons
					
			\item \textbf{Output}: whether each gene is a proper/quasi/non-transposon
		}
	\end{enumerate}
\end{algorithm}

\section{Linear sequences with duplicated genes}
In Scenario 3, consider $m$ linear gene sequences, where each sequence contains different numbers of copies of $n$ genes $1,\ldots,n$. We need to find the longest common subsequence. Here we only consider common subsequences that consist of all or none copies of the same gene, and the subsequence length is calculated by genes, not gene copies.

\subsection{A graph representation of the problem}
Similar to Scenario 1, we construct an auxiliary graph $\mathcal{G}$, where each vertex is a gene (not a copy of a gene). However, in this case, the auxiliary graph is undirected: There is an undirected edge between gene $g_i$ and gene $g_j$ if and only if all the copies of $g_i$ and $g_j$ keep their relative locations in all sequences. For example, consider two sequences $(1,2,3,2,3,4,5)$ and $(2,1,3,3,2,4,5)$. For gene pair $1,3$, the corresponding sequences are $(1,3,3)$ and $(1,3,3)$, meaning that there is an edge between $1$ and $3$. For gene pair $1,2$, the corresponding sequences are $(1,2,2)$ and $(2,1,2)$, meaning that there is no edge between $1$ and $2$. See Fig.~\ref{sce3} for the auxiliary graph in this case. 

\begin{figure}[htb]
	\begin{center}
		$\xymatrix{
			& 1\ar@{-}[dr]\ar@{-}[ddr]\ar@{-}[ddl] &\\
			2\ar@{-}[d]\ar@{-}[drr] &     & 3\ar@{-}[dll]\ar@{-}[d]   \\
			4\ar@{-}[rr]     &     & 5       }$
	\end{center}
	\caption{The auxiliary graph $\mathcal{G}$ of two sequences $(1,2,3,2,3,4,5)$ and $(2,1,3,3,2,4,5)$. The unique largest complete subgraph is $\{1,3,4,5\}$, meaning that the unique longest common sequence is $(1,3,3,4,5)$. Thus $1,3,4,5$ are non-transposons, and $2$ is a proper-transposon.}
	\label{sce3}
\end{figure}

\begin{definition}
	A subgraph of $\mathcal{G}$ consists of some genes $g_1,\ldots,g_l$ and the edges between them. In a subgraph, if there is an edge between any two genes, this subgraph is called a complete subgraph (also called a clique).
\end{definition}

If all copies of genes $g_1,\ldots,g_l$ keep their relative locations in all linear sequences, we say that $g_1,\ldots,g_l$ form a common subsequence. In this case, there is an edge in $\mathcal{G}$ between any two genes in $g_1,\ldots,g_l$, meaning that they form a complete subgraph. The following Lemma~\ref{l2} shows that the inverse also holds. Therefore, there is a bijection between common subsequences and complete subgraphs. \emph{The problem of determining the longest common subsequence now becomes determining the largest complete subgraph of $\mathcal{G}$.} 
\begin{lemma}
	In Scenario 3, if $g_1,\ldots,g_k$ form a complete subgraph in $\mathcal{G}$, then $g_1,\ldots,g_k$ form a common subsequence.
	\label{l2}
\end{lemma}
\begin{proof}
	Only consider copies of $g_1,\ldots,g_k$ in these sequences. If $g_1,\ldots,g_k$ do not form a common subsequence, find the first digit that such sequences differ. Assume $g_p$ and $g_q$ can both appear in this digit. Then $g_p,g_q$ cannot form a common subsequence, and there is no edge between $g_p$ and $g_q$. 
	
	We illustrate this proof with Fig.~\ref{sce3}: For genes $2,3,4$, the sequences are $(2,3,2,3,4)$ and $(2,3,3,2,4)$. The third digit is different, where $2$ and $3$ can both appear. Then the sequences for genes $2,3$, $(2,3,2,3)$ and $(2,3,3,2)$, cannot match, and there is no edge between $2$ and $3$.
\end{proof}

\subsection{A heuristic algorithm}
The above discussion shows that given gene sequences, we can construct an undirected graph $\mathcal{G}$, so that there is a bijection between common subsequences and complete subgraphs. The inverse also holds: We can construct corresponding gene sequences for a graph. 

\begin{lemma}
	Given an undirected graph $\mathcal{G}$, we can construct two gene sequences, so that there is a bijection between common subsequences and complete subgraphs.
	\label{le}
\end{lemma}
\begin{proof}
	Assume the graph has $n$ genes. We start with two sequences $(1,2,\ldots,n)$ and$(1,2,\ldots,n)$. For each pair of genes $g_i,g_j$, if there is no edge between them in $\mathcal{G}$, add $g_i,g_j$ to the end of the first sequence, and $g_j,g_i$ to the end of the second sequence. Then $g_i,g_j$ cannot both appear in a common subsequence, and this operation does not affect other gene pairs.
	
	For example, corresponding to Fig.~\ref{sce3}, we start with $(1,2,3,4,5)$ and $(1,2,3,4,5)$. Since there is no edge between $1,2$, we add them to have $(1,2,3,4,5,1,2)$ and $(1,2,3,4,5,2,1)$. Since there is no edge between $2,3$, we add them to have $(1,2,3,4,5,1,2,2,3)$ and $(1,2,3,4,5,2,1,3,2)$. These two sequences corresponds to Fig.~\ref{sce3}.
\end{proof}

Combining Lemma~\ref{l2} and Lemma~\ref{le}, we obtain the following result:
\begin{proposition}
	Finding the longest common sequence in Scenario 3 is NP-hard.
	\label{p1}
\end{proposition}
\begin{proof}
	For an undirected graph, we can use Lemma~\ref{le} to construct corresponding sequences. If we have the solution of finding the longest common sequence in Scenario 3, then we can find the largest complete subgraph in an extra polynomial time.
	
	For gene sequences in Scenario 3, we can construct corresponding auxiliary graph. If we have the solution of finding the largest complete subgraph, then we can use Lemma~\ref{l2} to find the longest common sequence in Scenario 3 in an extra polynomial time.
	
	Therefore, finding the longest common sequence in Scenario 3 and finding the largest complete subgraph are equivalent.  The problem of determining the largest complete subgraph is just the maximum clique problem, which is NP-hard \cite{valiente2002algorithms}. Thus finding the longest common sequence in Scenario 3 is also NP-hard. This means it is not likely to design an algorithm that always correctly determines the longest common subsequence in polynomial time. 
\end{proof}

We still determine transposons by finding the largest complete subgraph in $\mathcal{G}$, and we can design a greedy heuristic algorithm that only fails in rare cases. Readers may refer to a review for more details about the maximum clique problem \cite{wu2015review}.

\begin{definition}
	In graph $\mathcal{G}$, the degree of a gene $g$ is the number of edges linking $g$. In a complete graph of $p$ genes, where any two genes have an edge in between, each gene has degree $p-1$. 
\end{definition}

The idea is simple: In the auxiliary graph $\mathcal{G}$, repeatedly abandon the gene with the smallest degree (and also edges linking this gene) until the remaining genes form a complete subgraph. See Algorithm~\ref{alg5} for the details of this greedy heuristic method. 

\begin{algorithm}[!htbp]
	\caption{A heuristic method for detecting transposons in Scenario 3.}
	\label{alg5}
	\vspace{-\bigskipamount}
	\ \\
	\begin{enumerate}
		{	\item \textbf{Input}
			
			\quad $m$ linear sequences of genes $1,\ldots,n$, where each gene can have multiple copies
			
			\item \textbf{Construct} the auxiliary graph $\mathcal{G}$:
			
			\quad Vertices of $\mathcal{G}$ are all the genes $1,\ldots,n$ (not their copies)
			
			\quad \textbf{For} each pair of genes $g_i,g_j$
			
			\quad\quad   \textbf{If} all copies of $g_i$ and $g_j$ keep their relative locations in all $m$ sequences
			
			\quad\quad\quad \textbf{Add} an undirected edge between $g_i$ and $g_j$ in $\mathcal{G}$
			
			\quad\quad\textbf{End} of if
			
			\quad\textbf{End} of for
			
			\textbf{Calculate} the degree for each gene in $\mathcal{G}$
			
			\item \textbf{While} true
			
			\quad\textbf{Find} a gene $g_i$ with the smallest degree $d_i$ in $\mathcal{G}$ 
			
			\quad \% If the minimal $g_i$ is not unique, choose one randomly
			
			\quad \textbf{If} $d_i+1$ is smaller than the number of genes in $\mathcal{G}$
			
			\quad \quad \textbf{Delete} $g_i$ and edges linking $g_i$ in $\mathcal{G}$ 
			
			\quad \quad \textbf{Update} the degrees of other genes
			
			\quad \textbf{Else}
			
			\quad \% The remaining genes form a complete subgraph
			
			\quad \quad \textbf{Break} the while loop
			
			\quad\textbf{End} of if
			
			\textbf{End} of while
			
			\% The final $\mathcal{G}$ is a complete subgraph of the original $\mathcal{G}$, and it is likely to be the largest one
			
			\item \textbf{Output} genes in the final $\mathcal{G}$ are not transposons, and genes not in the final $\mathcal{G}$ are transposons
		}
	\end{enumerate}
\end{algorithm}

We test Algorithm~\ref{alg5} on random graphs. Construct a random graph with $n$ genes, and any two genes have probability $0.5$ to have an edge in between. Use brute-force search to find the maximum clique, and compare its size with the result of Algorithm~\ref{alg5}. For each $n\le 15$, we repeat this for $10000$ times, and every time Algorithm~\ref{alg5} returns the correct result. Therefore, we can claim that Algorithm~\ref{alg5} is a good heuristic algorithm that fails with a very small probability. Since finding the true maximum clique requires exponentially slow brute-force search, we do not test on very large graphs.

Nevertheless, Algorithm~\ref{alg5} does not always produce the correct result. See Fig.~\ref{ce2} for a counterexample. Here genes $1,2,3,4,5,6$ have degree $4$, while genes $7,8,9,10$ have degree $3$. When applying Algorithm~\ref{alg5}, genes $7,8,9,10$ are first abandoned, and the final result just has three genes, such as $1,3,5$. However, the largest complete graph is $7,8,9,10$. Besides, Algorithm~\ref{alg5} can only determine one (possibly longest) common subsequence. Thus we cannot determine the existence of quasi-transposons.

\begin{figure}[htb]
	\begin{center}
		$\xymatrix{
			3\ar@{-}[d]\ar@{-}[dr]\ar@{-}[drr]\ar@{-}[r]&1\ar@{-}[dl]\ar@{-}[r]\ar@{-}[dr]&4\ar@{-}[dll]\ar@{-}[dl]\ar@{-}[d]&7\ar@{-}[d]\ar@{-}[r]\ar@{-}[dr]&8\ar@{-}[d]\ar@{-}[dl]\\
			5\ar@{-}[r]&2\ar@{-}[r]&6&10\ar@{-}[r]&9
		}$
	\end{center}
	\caption{The auxiliary graph $\mathcal{G}$ of linear sequences $(7,8,9,10,1,1,2,3,3,4,5,5,6)$ and $(1,2,1,3,4,3,5,6,5,7,8,9,10)$. This counterexample fails Algorithm~\ref{alg5}.}
	\label{ce2}
\end{figure}


Assume we have $m$ sequences with $n$ genes. In general, the copy number of a gene is small, and we can assume the length of each sequence is $\mathcal{O}(n)$. The time complexities of Step 2 and Step 3 in Algorithm~\ref{alg5} are $\mathcal{O}(mn^2)$ and $\mathcal{O}(n^2)$, and the overall time complexity is $\mathcal{O}(mn^2)$. The space complexity is trivially $\mathcal{O}(mn+n^2)$. 

%
%
%
%
%
%
%
%
%
%
%
%
%

\section{Circular sequences with duplicated genes}
In Scenario 4, consider $m$ circular gene sequences, where each sequence contains different numbers of copies of $n$ genes $1,\ldots,n$. We need to find the longest common subsequence. Here we only consider common subsequences that consist of all or none copies of the same gene, and the subsequence length is calculated by genes, not gene copies.

In this scenario, Lemma~\ref{l2} does not hold. For example, we can consider a circular sequence $(1,2,3)$ and its mirror symmetry. These two sequences are different, but any two genes form a common subsequence. However, inspired by Lemma~\ref{l2}, we have the following conjecture. 

\begin{conjecture}
	In Scenario 4, if any three genes $g_i,g_j,g_l$ in $g_1,\ldots,g_k$ form a common subsequence, then $g_1,\ldots,g_k$ form a common subsequence. 
	\label{c2}
\end{conjecture}

For now, we do not know if Conjecture~\ref{c2} is correct or not. If it is valid, then we can try to find a group of genes, where any three of them form a common subsequence. This group of genes forms a common subsequence.

Construct a $3$-uniform hypergraph $\mathcal{G}$ as following \cite{Diestel}: vertices are genes $1,\ldots,n$; there is a $3$-hyperedge (undirected) that links genes $g_i,g_j,g_k$ if and only if they form a common subsequence. The longest common subsequence corresponds to the largest complete subgraph (any three genes are linked by a $3$-hyperedge). The maximum clique problem for $3$-uniform hypergraphs is NP-hard \cite{wu2015review}. Similar to the proof of Proposition~\ref{p1}, we can prove that finding the longest common subsequence in Scenario 4 is also NP-hard.

We have a simple idea: Repeatedly delete the gene that has the smallest degree, until we have a complete subgraph that any three genes have a $3$-hyperedge that links them. We summarize this greedy heuristic method as Algorithm~\ref{alg7}. 

We test Algorithm~\ref{alg7} on random graphs. Construct a random graph with $n$ genes, and any two genes have probability $0.5$ to have an edge in between. Use brute-force search to find the maximum clique, and compare its size with the result of Algorithm~\ref{alg7}. For each $n\le 15$, we repeat this for $10000$ times, and every time Algorithm~\ref{alg7} returns the correct result. Therefore, we can claim that Algorithm~\ref{alg7} is a good heuristic algorithm that fails with a very small probability. Since finding the true maximum clique requires exponentially slow brute-force search, we do not test on very large graphs.

Nevertheless, Algorithm~\ref{alg7} does not always produce the correct result. See Fig.~\ref{ce3} for a counterexample. Here each gene in $1,2,3,4,5,6$ has degree $4$, while each gene in $7,8,9,10$ has degree $3$ .When applying Algorithm~\ref{alg7}, genes $7,8,9,10$ are first deleted, and the final result just has three genes, such as $(1,3,5)$. However, the longest common subsequence $(7,8,9,10)$ has four genes.

Assume we have $m$ sequences with $n$ genes. In general, the copy number of a gene is small, and we can assume the length of each sequence is $\mathcal{O}(n)$. The time complexities of Step 2 and Step 3 in Algorithm~\ref{alg7} are $\mathcal{O}(mn^3)$ and $\mathcal{O}(n^3)$, and the overall time complexity is $\mathcal{O}(mn^3)$. The space complexity is trivially $\mathcal{O}(mn+n^3)$. 

\begin{figure}[htb]
	\begin{center}
		$\xymatrix{
			1\ar@{-}[r]&2\ar@{-}[r]&7\ar@{-}[r]&3\ar@{-}[r]&4\ar@{-}[d]&2\ar@{-}[r]&1\ar@{-}[r]&10\ar@{-}[r]&4\ar@{-}[r]&3\ar@{-}[d]\\
			10\ar@{-}[u]&6\ar@{-}[l]&5\ar@{-}[l]&9\ar@{-}[l]&8\ar@{-}[l]&9\ar@{-}[u]&5\ar@{-}[l]&6\ar@{-}[l]&8\ar@{-}[l]&7\ar@{-}[l]\\
			1\ar@{-}[r]&2\ar@{-}[r]&9\ar@{-}[r]&3\ar@{-}[r]&4\ar@{-}[d]&2\ar@{-}[r]&1\ar@{-}[r]&8\ar@{-}[r]&4\ar@{-}[r]&3\ar@{-}[d]\\
			8\ar@{-}[u]&6\ar@{-}[l]&5\ar@{-}[l]&7\ar@{-}[l]&10\ar@{-}[l]&7\ar@{-}[u]&5\ar@{-}[l]&6\ar@{-}[l]&10\ar@{-}[l]&9\ar@{-}[l]
		}$
	\end{center}
	\caption{Four circular sequences. The longest common subsequence is $(7,8,9,10)$. This counterexample fails Algorithm~\ref{alg7}.}
	\label{ce3}
\end{figure}

\begin{algorithm}[!htbp]
	\caption{A heuristic method for detecting transposons in Scenario 4.}
	\label{alg7}
	\vspace{-\bigskipamount}
	\ \\
	\begin{enumerate}
		{	\item \textbf{Input}
			
			\quad $m$ circular sequences of genes $1,\ldots,n$, where each gene can have multiple copies
			
			\item \textbf{Construct} the auxiliary graph $\mathcal{G}$:
			
			\quad Vertices of $\mathcal{G}$ are all the genes $1,\ldots,n$ (not their copies)
			
			\quad \textbf{For} each gene triple $g_i,g_j,g_k$
			
			\quad\quad   \textbf{If} all copies of $g_i,g_j,g_k$ keep their relative locations in all $m$ sequences
			
			\quad\quad\quad \textbf{Add} a $3$-hyperedge that links $g_i,g_j,g_k$ in $\mathcal{G}$
			
			\quad\quad\textbf{End} of if
			
			\quad\textbf{End} of for			
			
			\item \textbf{While} there exist three genes that do not share a $3$-hyperedge
			
			\quad \textbf{Calculate} the degree for each gene in $\mathcal{G}$
			
			\quad\textbf{Delete} the gene with the smallest degree and $3$-hyperedges that links this gene
			
			\quad \% If there are multiple genes with the smallest degree, delete one randomly
						
			\textbf{End} of while
			
			\% After this while loop, any three genes form a common subsequence
			
			\% If Conjecture~\ref{c2} holds, the remaining genes form a common subsequence
			
			\item \textbf{Output} remaining genes are not transposons, and other genes are transposons
		}
	\end{enumerate}
\end{algorithm}

\section{Discussion}
In this paper, we study the problem of determining transposons in gene sequences. Depending on whether the gene sequences are linear or circular, and whether genes have multiple copies, we classify the problem into four scenarios. We first transform the problems of determining transposons into finding the longest common subsequences, and then transform them into graph theory problems. For the first two scenarios without duplicated genes, we develop complete algorithms with polynomial complexities. For the latter two scenarios with duplicated genes, the problems are NP-hard, and we develop fast algorithms that only fail in rare cases. We also propose some unresolved conjectures in discrete mathematics.

In Scenario 1 and Scenario 2 (linear/circular sequences without duplicated genes), if each sequence has $n$ genes, and the longest common subsequence has length $n-k$, then there are at most $k$ proper-transposons. About quasi-transposons, inspired by Lemma~\ref{l1}, we have the following guess.

\begin{conjecture}
	Consider $m$ linear/circular sequences with $n$ genes without multiple copies. Assume the length of the longest common subsequence is $n-k$, and there are $l$ proper-transposons. Then the number of quasi-transposons is no larger than $2(k-l)$.
	\label{c3}
\end{conjecture}

When $l+2(k-l)\le n$, in both linear and circular scenarios, we can find examples with $2(k-l)$ quasi-transposons.

When transposons have been determined, we can use them to compare the genomes of different species, and such comparisons can be combined with other measurements between species, such as metrics on developmental trees \cite{wang2022two}. Such comparisons can be also extended to different tissues to help with the prediction of tissue transplantation experiments \cite{wang2021inference}. Besides, for some species, cells at different positions have different gene expression patterns, which might be related to transposons \cite{wang2020biological}. 

A gene $g_i$ might be missing in some sequences. Since $g_i$ is not in any longest common subsequence, it should be a proper-transposon. This gene can be directly removed before applying corresponding algorithms.

We can adopt a stricter definition of transposons to exclude a gene which only changes its relative position in a few (no more than $l$, where $l$ is small enough) sequences. Then we should consider the longest sequence which is a common subsequence of at least $m-l$ sequences. We can run the corresponding algorithm for every $m-l$ sequences. Thus the total time complexity will be multiplied by a factor of $m^l$.

\section*{Acknowledgements}
This research was partially supported by NIH grant R01HL146552. The author would like to thank Zhongkai Zhao for helping with designing Algorithm~\ref{alg1}. The author would like to thank Lucas B\"ottcher and an anonymous reviewer for providing helpful comments.

\bibliographystyle{vancouver}
\bibliography{Transposons}

\begin{thebibliography}{10}

\bibitem{ivics2010expanding}
Ivics Z, Izsv{\'a}k Z.
\newblock The expanding universe of transposon technologies for gene and cell
  engineering.
\newblock Mob DNA. 2010;1(1):1--15.

\bibitem{mills2007transposable}
Mills RE, Bennett EA, Iskow RC, Devine SE.
\newblock Which transposable elements are active in the human genome?
\newblock Trends Genet. 2007;23(4):183--191.

\bibitem{zhou2020dna}
Zhou W, Liang G, Molloy PL, Jones PA.
\newblock DNA methylation enables transposable element-driven genome expansion.
\newblock Proc Natl Acad Sci USA. 2020;117(32):19359--19366.

\bibitem{denicola2015utility}
DeNicola GM, Karreth FA, Adams DJ, Wong CC.
\newblock The utility of transposon mutagenesis for cancer studies in the era
  of genome editing.
\newblock Genome Biol. 2015;16(1):1--15.

\bibitem{kazazian1988haemophilia}
Kazazian HH, Wong C, Youssoufian H, Scott AF, Phillips DG, Antonarakis SE.
\newblock {Haemophilia A} resulting from de novo insertion of {L1} sequences
  represents a novel mechanism for mutation in man.
\newblock Nature. 1988;332(6160):164--166.

\bibitem{mustajoki1999insertion}
Mustajoki S, Ahola H, Mustajoki P, Kauppinen R.
\newblock Insertion of {Alu} element responsible for acute intermittent
  porphyria.
\newblock Hum Mutat. 1999;13(6):431--438.

\bibitem{zhou2014multi}
Zhou D, Wang Y, Wu B.
\newblock A multi-phenotypic cancer model with cell plasticity.
\newblock J Theor Biol. 2014;357:35--45.

\bibitem{niu2015phenotypic}
Niu Y, Wang Y, Zhou D.
\newblock The phenotypic equilibrium of cancer cells: {From} average-level
  stability to path-wise convergence.
\newblock J Theor Biol. 2015;386:7--17.

\bibitem{niu2019transposable}
Niu XM, Xu YC, Li ZW, Bian YT, Hou XH, Chen JF, et~al.
\newblock Transposable elements drive rapid phenotypic variation in {Capsella}
  rubella.
\newblock Proc Natl Acad Sci USA. 2019;116(14):6908--6913.

\bibitem{chen2016overshoot}
Chen X, Wang Y, Feng T, Yi M, Zhang X, Zhou D.
\newblock The overshoot and phenotypic equilibrium in characterizing cancer
  dynamics of reversible phenotypic plasticity.
\newblock J Theor Biol. 2016;390:40--49.

\bibitem{jiang2017phenotypic}
Jiang DQ, Wang Y, Zhou D.
\newblock Phenotypic equilibrium as probabilistic convergence in
  multi-phenotype cell population dynamics.
\newblock PLOS ONE. 2017;12(2):e0170916.

\bibitem{noorani2020crispr}
Noorani I, Bradley A, de~la Rosa J.
\newblock {CRISPR} and transposon in vivo screens for cancer drivers and
  therapeutic targets.
\newblock Genome Biol. 2020;21(1):1--22.

\bibitem{angelini2022model}
Angelini E, Wang Y, Zhou JX, Qian H, Huang S.
\newblock A model for the intrinsic limit of cancer therapy: {Duality} of
  treatment-induced cell death and treatment-induced stemness.
\newblock PLOS Comput Biol. 2022;18(7):e1010319.

\bibitem{nowacki2009functional}
Nowacki M, Higgins BP, Maquilan GM, Swart EC, Doak TG, Landweber LF.
\newblock A functional role for transposases in a large eukaryotic genome.
\newblock Science. 2009;324(5929):935--938.

\bibitem{babakhani2018transposons}
Babakhani S, Oloomi M.
\newblock Transposons: the agents of antibiotic resistance in bacteria.
\newblock J Basic Microbiol. 2018;58(11):905--917.

\bibitem{rahrmann2009identification}
Rahrmann EP, Collier LS, Knutson TP, Doyal ME, Kuslak SL, Green LE, et~al.
\newblock Identification of {PDE4D} as a proliferation promoting factor in
  prostate cancer using a {Sleeping Beauty} transposon-based somatic
  mutagenesis screen.
\newblock Cancer Res. 2009;69(10):4388--4397.

\bibitem{xia2020pde}
Xia M, Greenman CD, Chou T.
\newblock {PDE} models of adder mechanisms in cellular proliferation.
\newblock SIAM J Appl Math. 2020;80(3):1307--1335.

\bibitem{dessalles2021naive}
Dessalles R, Pan Y, Xia M, Maestrini D, D’Orsogna MR, Chou T.
\newblock How Naive T-Cell Clone Counts Are Shaped By Heterogeneous Thymic
  Output and Homeostatic Proliferation.
\newblock Front Immunol. 2021;12.

\bibitem{wang2022inference}
Wang Y, Wang Z.
\newblock Inference on the structure of gene regulatory networks.
\newblock J Theor Biol. 2022;539:111055.

\bibitem{sha2020inference}
Sha Y, Wang S, Zhou P, Nie Q.
\newblock Inference and multiscale model of epithelial-to-mesenchymal
  transition via single-cell transcriptomic data.
\newblock Nucleic Acids Res. 2020;48(17):9505--9520.

\bibitem{zhou2021dissecting}
Zhou P, Wang S, Li T, Nie Q.
\newblock Dissecting transition cells from single-cell transcriptome data
  through multiscale stochastic dynamics.
\newblock Nat Commun. 2021;12(1):1--15.

\bibitem{gardner2017mobile}
Gardner EJ, Lam VK, Harris DN, Chuang NT, Scott EC, Pittard WS, et~al.
\newblock {The Mobile Element Locator Tool} ({MELT}): population-scale mobile
  element discovery and biology.
\newblock Genome Res. 2017;27(11):1916--1929.

\bibitem{chen2019ervcaller}
Chen X, Li D.
\newblock {ERVcaller}: identifying polymorphic endogenous retrovirus and other
  transposable element insertions using whole-genome sequencing data.
\newblock Bioinformatics. 2019;35(20):3913--3922.

\bibitem{yu2021benchmark}
Yu T, Huang X, Dou S, Tang X, Luo S, Theurkauf WE, et~al.
\newblock A benchmark and an algorithm for detecting germline transposon
  insertions and measuring de novo transposon insertion frequencies.
\newblock Nucleic Acids Res. 2021;49(8):e44--e44.

\bibitem{orozco2020measuring}
Orozco-Arias S, Pi{\~n}a JS, Tabares-Soto R, Castillo-Ossa LF, Guyot R, Isaza
  G.
\newblock Measuring performance metrics of machine learning algorithms for
  detecting and classifying transposable elements.
\newblock Processes. 2020;8(6):638.

\bibitem{goubert2022beginner}
Goubert C, Craig RJ, Bilat AF, Peona V, Vogan AA, Protasio AV.
\newblock A beginner’s guide to manual curation of transposable elements.
\newblock Mob DNA. 2022;13(1):1--19.

\bibitem{evrony2021applications}
Evrony GD, Hinch AG, Luo C.
\newblock Applications of single-cell DNA sequencing.
\newblock Annu Rev Genomics Hum Genet. 2021;22:171.

\bibitem{lin2011changes}
Lin CH, Lian CY, Hsiung CA, Chen FC; BioMed Central.
\newblock Changes in transcriptional orientation are associated with increases
  in evolutionary rates of enterobacterial genes.
\newblock BMC Bioinform. 2011;12(9):1--8.

\bibitem{chen2004space}
Chen ZZ, Gao Y, Lin G, Niewiadomski R, Wang Y, Wu J.
\newblock A space-efficient algorithm for sequence alignment with inversions
  and reversals.
\newblock Theor Comput Sci. 2004;325(3):361--372.

\bibitem{imbeault2017krab}
Imbeault M, Helleboid PY, Trono D.
\newblock KRAB zinc-finger proteins contribute to the evolution of gene
  regulatory networks.
\newblock Nature. 2017;543(7646):550--554.

\bibitem{zimin2017hybrid}
Zimin AV, Puiu D, Luo MC, Zhu T, Koren S, Mar{\c{c}}ais G, et~al.
\newblock Hybrid assembly of the large and highly repetitive genome of Aegilops
  tauschii, a progenitor of bread wheat, with the MaSuRCA mega-reads algorithm.
\newblock Genome Res. 2017;27(5):787--792.

\bibitem{reneker2012long}
Reneker J, Lyons E, Conant GC, Pires JC, Freeling M, Shyu CR, et~al.
\newblock Long identical multispecies elements in plant and animal genomes.
\newblock Proc Natl Acad Sci USA. 2012;109(19):E1183--E1191.

\bibitem{diop2020pseudomolecule}
Diop SI, Subotic O, Giraldo-Fonseca A, Waller M, Kirbis A, Neubauer A, et~al.
\newblock A pseudomolecule-scale genome assembly of the liverwort Marchantia
  polymorpha.
\newblock Plant J. 2020;101(6):1378--1396.

\bibitem{hirschberg1975linear}
Hirschberg DS.
\newblock A linear space algorithm for computing maximal common subsequences.
\newblock Commun ACM. 1975;18(6):341--343.

\bibitem{backurs2015edit}
Backurs A, Indyk P.
\newblock Edit distance cannot be computed in strongly subquadratic time
  (unless {SETH} is false).
\newblock In: {Proceedings of the Forty-Seventh Annual ACM Symposium on Theory
  of Computing}; 2015. p. 51--58.

\bibitem{kiyomi2021longest}
Kiyomi M, Horiyama T, Otachi Y.
\newblock Longest common subsequence in sublinear space.
\newblock Inf Process Lett. 2021;168:106084.

\bibitem{maier1978complexity}
Maier D.
\newblock The complexity of some problems on subsequences and supersequences.
\newblock J ACM. 1978;25(2):322--336.

\bibitem{blum2021solving}
Blum C, Djukanovic M, Santini A, Jiang H, Li CM, Many{\`a} F, et~al.
\newblock Solving longest common subsequence problems via a transformation to
  the maximum clique problem.
\newblock Comput Oper Res. 2021;125:105089.

\bibitem{wang2010fast}
Wang Q, Korkin D, Shang Y.
\newblock A fast multiple longest common subsequence ({MLCS}) algorithm.
\newblock IEEE Trans Knowl Data Eng. 2010;23(3):321--334.

\bibitem{mousavi2012improved}
Mousavi SR, Tabataba F.
\newblock An improved algorithm for the longest common subsequence problem.
\newblock Comput Oper Res. 2012;39(3):512--520.

\bibitem{islam2019chemical}
Islam M, Saifullah C, Asha ZT, Ahamed R, et~al.
\newblock Chemical reaction optimization for solving longest common subsequence
  problem for multiple string.
\newblock Soft Comput. 2019;23(14):5485--5509.

\bibitem{bergroth2000survey}
Bergroth L, Hakonen H, Raita T.
\newblock A survey of longest common subsequence algorithms.
\newblock In: Proceedings Seventh International Symposium on String Processing
  and Information Retrieval. SPIRE 2000. IEEE; 2000. p. 39--48.

\bibitem{huang2004fast}
Huang K, Yang CB, Tseng KT, et~al.
\newblock Fast algorithms for finding the common subsequences of multiple
  sequences.
\newblock In: {Proceedings of the International Computer Symposium}. Citeseer;
  2004. p. 1006--1011.

\bibitem{wei2020path}
Wei S, Wang Y, Yang Y, Liu S.
\newblock A path recorder algorithm for {Multiple Longest Common Subsequences
  (MLCS)} problems.
\newblock Bioinformatics. 2020;36(10):3035--3042.

\bibitem{wu2015review}
Wu Q, Hao JK.
\newblock A review on algorithms for maximum clique problems.
\newblock Eur J Oper Res. 2015;242(3):693--709.

\bibitem{kang2014flexibility}
Kang Y, Gu C, Yuan L, Wang Y, Zhu Y, Li X, et~al.
\newblock Flexibility and symmetry of prokaryotic genome rearrangement reveal
  lineage-associated core-gene-defined genome organizational frameworks.
\newblock mBio. 2014;5(6):e01867--14.

\bibitem{rowley2018organizational}
Rowley MJ, Corces VG.
\newblock Organizational principles of {3D} genome architecture.
\newblock Nat Rev Genet. 2018;19(12):789--800.

\bibitem{wang2020causal}
Wang Y, Wang L.
\newblock Causal inference in degenerate systems: {An} impossibility result.
\newblock In: {International Conference on Artificial Intelligence and
  Statistics}. PMLR; 2020. p. 3383--3392.

\bibitem{verma2019architecture}
Verma SC, Qian Z, Adhya SL.
\newblock Architecture of the {Escherichia coli} nucleoid.
\newblock PLOS Genet. 2019;15(12):e1008456.

\bibitem{ibal2019information}
Ibal JC, Pham HQ, Park CE, Shin JH.
\newblock Information about variations in multiple copies of bacterial 16S
  {rRNA} genes may aid in species identification.
\newblock PLOS ONE. 2019;14(2):e0212090.

\bibitem{wang2018some}
Wang Y.
\newblock Some Problems in Stochastic Dynamics and Statistical Analysis of
  Single-Cell Biology of Cancer [Ph.D. thesis].
\newblock University of Washington; 2018.

\bibitem{valiente2002algorithms}
Valiente G.
\newblock Algorithms on Trees and Graphs.
\newblock Berlin: Springer; 2002.

\bibitem{Diestel}
Diestel R.
\newblock Graph Theory.
\newblock 5th ed. Berlin: Springer; 2017.

\bibitem{wang2022two}
Wang Y.
\newblock Two metrics on rooted unordered trees with labels.
\newblock Algorithms Mol Biol. 2022;17(1):1--17.

\bibitem{wang2021inference}
Wang Y, Zhang B, Kropp J, Morozova N.
\newblock Inference on tissue transplantation experiments.
\newblock J Theor Biol. 2021;520:110645.

\bibitem{wang2020biological}
Wang Y, Kropp J, Morozova N.
\newblock Biological notion of positional information/value in morphogenesis
  theory.
\newblock Int J Dev Biol. 2020;64(10-11-12):453--463.

\end{thebibliography}

\end{document}